\documentclass[twoside,a4paper,11pt]{article}
\usepackage{color}  
\usepackage{amsfonts}                   
\usepackage[psamsfonts]{amssymb}
\usepackage[bottom]{footmisc}
\usepackage{amsmath}
\usepackage{multicol}
\usepackage{a4wide}                     
\usepackage{fancyhdr}

\newcounter{eg}                         \newtheorem{eg}{Example}[section]        
\def\beg{\begin{eg}\rm}                 \def\eeg{\hfill\sq\end{eg}}                

\newcommand{\initiate}{\setcounter{equation}{0}} 
\def\b#1{{\mathbb #1}}

\def\c#1{{\cal #1}}

\def\Dirac{{\raise0.09em\hbox{/}}\kern-0.69em D}

\def\exterior{{{\raise0.2em\hbox{$\scriptstyle\bigwedge$}}{}}}

\def\kb{i\kbar}

\def\kbar{{\mathchar'26\mkern-9muk}}

\def\lesssim{\mathrel{\hbox{\rlap{hbox{\lower8pt\hbox{$\sim$}}}\hbox{$<$}}}}

\def\sq{\hbox{\rlap{$\sqcap$}$\sqcup$}}

\def\p{\partial}

\def\dfrac #1#2{\displaystyle{\frac{#1}{#2}}}

\def\k{\kern-.1em\mathbin{,}\kern-.1em}
 
\def\hk{\kern.12em\raise-1em\hbox{$\hat{\raise1em\hbox{,}}$}\kern.12em}

\begin{document}

\title{Noncommutative de~Sitter and FRW spaces.}
\author{Maja Buri\'c $^{1}$\thanks{majab@ipb.ac.rs} \ 
                   and
        John Madore $^{2}$\thanks{madore@th.u-psud.fr} 
                   \\[15pt]$\strut^{1}$
        University of Belgrade,  Faculty of Physics, P.O. Box 44
                   \\
        SR-11001 Belgrade 
                   \\[5pt]$\strut^{2}$
        Laboratoire de Physique Th\'eorique 
                   \\
        F-91405 Orsay   
       }
 
\date{}
\maketitle
\vfill
\parindent 0pt

\begin{abstract}
  Several versions of fuzzy four-dimensional de~Sitter
space are constructed using the noncommutative 
frame formalism. Although all  noncommutative spacetimes
which are found  have commutative de~Sitter metric as 
a classical limit, the algebras 
and the differential calculi which define them have many 
differences which we derive and discuss.
\end{abstract}

\vfill
\setlength{\parskip}{15pt plus5pt minus0pt}
\setlength{\textheight}{24.0cm}

\pagestyle{plain}

\initiate
\section{Introduction}

Various and sundry reasons have been put forward to entice physicists
into spacetime noncommutativity. It has for example been argued that
noncommutative geometries could incorporate some aspects of the
quantized gravitational field.  There seems further to be no obvious
physical reason to extrapolate the commutativity of coordinates and
the corresponding description of spacetime as a manifold to
arbitrarily small length scales. However, if one approaches the task
of defining `noncommutative space' from a physical or physically
useful point of view and not as a purely mathematical abstraction,
there is a long list of properties which one might like to
incorporate in order to be able to use the standard language
of dynamics and symmetries. 

Inspired by the fact that the geometry of a smooth manifold can be
described in terms of the algebra of smooth functions defined on it,
probably the most plausible starting point is to define a
noncommutative space as an algebra ${\cal A}$ of linear operators.
This approach in many ways inherits intuition from quantum 
mechanics. There are other approaches which rest on insights 
and constructions from string theory or conformal field
theory, \cite{Heckman:2014xha,Freidel:2015pka,Blumenhagen:2013zpa}.
Apart from  spacetime, one always has field equations
 which involve usually the action of a Laplace or Dirac operator: 
one must that is
 define  derivations. It is not {\it a priori} obvious
which properties derivations should possess but a natural condition is the
Leibniz rule; there are however important models in which  differential
operators do not obey it~\cite{Aschieri:2005zs,Wess:2006cm,Chandra:2014qva}. 
 At the risk of narrowing down the class of possible structures, 
we attempt  to extend the various elements of classical geometry
to a simple and  well studied structure such as algebra ${\cal A}$. 
One  important physical aspect is symmetry:  
symmetries are in the algebraic framework represented quite naturally.  
And finally, a relevant question which one has to address
 is that of the commutative or 
classical limit. Given that associative algebras are rigid
structures constrained by the Jacobi identities, it is not 
clear whether it is possible to fulfil all these 
 requirements in a physically or mathematically satisfactory way.

There are  indications, both from quantum mechanics and general
relativity, that when introducing a noncommutative space 
one should consider not just spacetime that is position
space, but the full phase space.  In quantum
mechanics  phase space consists of commutative coordinates $x^\mu$
and (here taken antihermitian) 
momenta $\hbar p_\alpha= \delta^\mu_\alpha\, \p_\mu$. The adjoint 
action of momenta $p_\alpha$ on
elements $f(x)\in {\cal A}$ of the position algebra defines the 
derivations,
\begin{equation}
[p_\alpha,f]= (e_\alpha f ) =  \delta^\mu_\alpha \, (\p_\mu f),
\end{equation}
in particular
\begin{equation}
[p_\alpha,x^\mu]=  \delta^\mu_\alpha .
\end{equation}
This can be seen perhaps more clearly by using explicitly
the Hilbert space ${\cal H}$ on which the representation of
 the algebra acts:
derivations appear as momenta.
In a completely analogous way one can interpret the frame
derivations $e_\alpha$ defined in the Cartan frame 
description of geometry,
\begin{equation}
 e_\alpha = e^\mu_\alpha(x)\, \p_\mu ,
\end{equation}
as momenta conjugate  to coordinates, again assuming the adjoint 
action  on the functions of coordinates. The  canonical commutators 
in the gravitational field change,
\begin{equation}
[p_\alpha,x^\mu] = e_\alpha x^\mu =e^\mu_\alpha(x),
\end{equation}
and  characterize the  curved 
spacetime because  the  $e_\alpha$ do not in general commute,
\begin{equation}
[e_\alpha,e_\beta] = C^\gamma{}_{\alpha\beta}(x)\, e_\gamma .  \label{GR}
\end{equation}
The last relation can be written as a commutator of the momenta
\begin{equation}
[p_\alpha,p_\beta] = C^\gamma{}_{\alpha\beta}(x) \, p_\gamma   \label{*}
\end{equation}
which acts in the adjoint representation on an arbitrary
function $f(x)$. Quite naturally, noncommutativity in momentum 
space is equivalent to curvature in position space. 
By classical duality between position and momenta, noncommutativity 
in position space then is related to
curvature in momentum space:  we thus conjecture that curvature and
noncommutativity are two aspects of the same reality.

We mentioned that it is not easy to extend all physical 
and geometrical requirements to noncommutative space.
A particularly delicate question in noncommutative geometry is 
that of dimension. On a commutative manifold to describe a point 
we need $n$ real parameters; this integer is also the dimension 
of the tangent space as each vector $X$ can be expanded as
\begin{equation}
 X=X^\mu(x)\p_\mu        .                      \label{X}
\end{equation}
The set of derivations $\{ X\}$ is left $\c{A}$-module and the de~Rham
differential is uniquely defined. Phase space has $2n$-dimensions.

Counting dimensions is different on  noncommutative space and in
order to obtain some intuition we point out the differences which
appear in the simplest examples. Let us take the space $M_2$ of
$\, 2\times2$ (or  $M_n$ of $\, n\times n$) complex matrices. As
a linear space it has 4 complex, or 8 real dimensions.  The subspace  
of hermitian matrices has 4 real dimensions, and a suitable 
basis is for example given by the Pauli matrices and unity,
 $\, \{ \sigma^i, \b{I} \}$.  However if we consider the set $M_2$
 as an algebra,  we need only two 
$\sigma$-matrices to generate it, for example
$\sigma_x$ and $\sigma_y $, as  $\sigma_x^2=\b{I}$, 
$\sigma_x\sigma_y =i\sigma_z$.  Therefore the number  
of generators of the algebra is 2 and one might conclude that 
its dimension is 2 as well. In similar manner, infinite-dimensional 
linear  algebra of operators on the Hilbert space of quadratically 
integrable functions of one variable $x$ is generated by 
 two operators, $x$ and $\p_x$, or $a$ and  $a^\dagger$. 

 If  we consider, on the other hand, a set of inner 
derivations $\{ X \}$ on matrix space $M_2 $,
\begin{equation}
 X_pf =[p,f],            \qquad \     p\in M_2 ,        \label{[p,f]}
\end{equation}
 we see that it is
of dimension 3 (that is, $n^2-1$ in $M_n$) because $\, [\b{I},f]=0$.
On noncommutative space the set of derivations is no longer a 
left module, that is $fX$ is not a derivation along with $X$ because
 it does not satisfy the Leibniz rule. Therefore in order to 
determine dimension of the tangent space  we have to `count' 
 momenta $p$ as a linear space over the real or complex numbers.
Consequently,  dimension of the tangent space always  differs 
from dimension  (the number of generators) of 
 the algebra itself. We see thus that the notion of dimension
does not have a precise meaning and
speaking of it we usually try to relate it  to the 
commutative limit of a given noncommutative geometry.
A further reason is that we wish to interpret a specific
algebra as a noncommutative space independently of 
its representation, whereas linear dimension is
 related to  dimensionality of  representations.

The noncommutative frame formalism \cite{book} solves
 this problem in the following manner. On ${\cal A}$ a differential 
$d$  can be defined for all vector fields in analogy with
the commutative case,
\begin{equation}
df(X) = Xf.
\end{equation}
But as we have seen, the linear space of all vector fields is `too big': 
we can redefine $d$ by restricting it on a subset $\{ e_\alpha\}$ 
of the set of all derivations. Thereby we define the tangent space. 
Let $\{ \theta^\alpha\}$ be the set of 1-forms dual to $e_\alpha$, 
 $\, \theta^\alpha(e_\beta)=\delta^\alpha_\beta$.  Then a
restriction of  $d$ is defined by
\begin{equation}
 df = (e_\alpha f) \theta^\alpha.
\end{equation}
The set of 1-forms is a bimodule, that is along with
$\theta^\alpha$,  $f\theta^\alpha$ and $\theta^\alpha f$ 
are 1-forms. Duality is  equivalent to the `frame condition'
\begin{align}
[f,\theta^\alpha]=0      ,\qquad   \forall f\in {\cal A}.             \label{[]}
\end{align}

Vector fields can always be given as commutators (\ref{[p,f]}), 
but momenta $p$ may not belong to position algebra 
${\cal A}$. Coordinates and momenta together generate phase space.
A peculiar property now is that dimension of  phase 
space is in general not equal to  $2\times\,$dimension of 
spacetime.  This comes about beacuse of noncommutativity of
coordinates: for the Heisenberg algebra for example, 
that is the two-dimensional Moyal space where
\begin{equation}
[x,y]=i,
\end{equation}
we can take $p_x=iy$, $p_y=-ix$ and get
\begin{equation}
 [p_i,x^j] =\delta^j_i ,
\end{equation}
 the flat frame. In this case phase space is identical to 
position space. Similar situation we have for matrix
algebras $M_n$ because on $M_n$ all derivations are inner.
We thus see that in the noncommutative frame formalism 
by the choice of the set $\{ e_\alpha\}$
we effectively fix the dimension of
spacetime (defined as  dimension of its tangent space).
This choice is by no means unique and reflects the property 
that differential $d$ on an algebra is not unque either.

An important characteristic of the noncommutative frame formalism is that,
if momenta generate the same algebra as coordinates that is if
all derivations are inner, they must satisfy a quadratic relation
\begin{equation}
2P^{\alpha\beta}{}_{\gamma\delta} p_{\alpha} p_\beta 
- F^\beta{}_{\gamma\delta} p_\beta - K_{\beta\gamma} = 0       ,         \label{qr}
\end{equation}
where $P^{\alpha\beta}{}_{\gamma\delta} $,  $F^\beta{}_{\gamma\delta}$ 
and $ K_{\beta\gamma}$ are constants. This relation follows from 
stability of Equation~(\ref{[]}) under the differential,
\begin{align}
 df 
= - [\theta, f], \qquad \ \theta = - p_\alpha\theta^\alpha  ,  \label{d}
\end{align}
 and constraint $d^2=0$.
We stress the simplicity and importance of these equations as well as
their content. On the one hand they define differential calculi on an
arbitrary algebra in much the same way that the de~Rham calculus is
defined on a smooth manifold. On the other hand we see immediately
that the calculus is not unique; each choice of the momenta
consistent with~(\ref{qr}) gives a different $d$. 
Finally,~(\ref{[]}) allows one to interpret 1-forms $\theta^\alpha$
as the moving frame assuming that the   frame  components of 
the metric are constants, for example
\begin{equation}
 g^{\alpha\beta} = \eta^{\alpha\beta}.
\end{equation}
In the commutative case there are no restrictions
from the commutators or from the associativity of the product; momenta
are necessarily external and from~(\ref{GR}) we see that there is no
analog of~(\ref{qr}).

\section{Noncommutative de~Sitter space, I}

We saw in the previous section that the frame formalism 
gives a definition of differential general enough to 
include commutative manifolds, quantum-mechanical phase 
space and noncommutative matrix spaces. Its main constituent 
is the moving frame which naturally incorporates geometry.  
To see whether this formalism can indeed describe 
noncommutative gravity we proceed by examples which fulfil
 previously mentioned requirements and have a certain
relevance  in  physics. We discussed in previous 
papers~\cite{Buric:2014ika,Buric:2009zz,Buric:2008th} various 
 rotationally invariant noncommutative spaces. 
In this paper we give examples of algebras 
 with spherical symmetry which can 
be considered as fuzzy versions of cosmological metrics: 
de~Sitter and Friedmann-Robertson-Walker. We find several
versions of noncommutative four-dimensional de~Sitter space
which have different spectral and symmetry properties, but 
the same limiting classical metric.  Another common feature 
which they share is that one needs to make some kind of dimensional 
extension to obtain a smooth noncommutative space. 

Perhaps the most natural idea, when we think of constructing
 noncommutative de~Sitter spacetime, is to start from 
the Lie algebra of the de~Sitter group
itself. This idea was in some details
put forward in ~\cite{Gazeau:2006hj,Gazeau:2009mi}, as a 
generalisation of the fuzzy sphere construction~\cite{fs}:
we shall in this section analyse and develop it.
Let us shortly recall the fuzzy sphere.
The possibility to interpret the $SO(3)$ group generators
$x^i$, $i=1,2,3$,  as Cartesian coordinates on the 
sphere is based on two facts. The first is that operators $x^i$ 
in the irreducible representations satisfy the Casimir relation
\begin{equation} 
 \delta_{ij}\, x^i x^j = {\cal C} = {\rm const}.
\end{equation}
This relation is the same as one which defines embedding
 of the two-sphere in the three-dimensional euclidean space.
The second  fact which  ensures smoothness is a possibility 
to define differential calculus. The differential can be written 
in form of a noncommutative frame~\cite{fs}, with the momenta 
given by
\begin{equation}
 p_a =\frac{1}{i\kbar}\,\delta_{ai}\, x^i, \qquad a=1,2,3 .                           \label{fsp}
\end{equation}
To justify that (\ref{fsp}) gives spherical geometry
one can either calculate the coordinate components of the metric
and obtain the projector to the sphere,
\begin{equation}
 g^{ij} = e^i_a e^j_b \delta^{ab} = {\cal C}\delta^{ij} - x^j x^i,
\end{equation}
or calculate the scalar curvature and obtain a constant. One can
also verify that the generators of rotations are 
Killing vectors, in much the same way se as we later do 
for de~Sitter generators on fuzzy de~Sitter  space.
In addition, there is a well defined commutative limit:
 the polynomial expansion of an arbitrary matrix  $f\in M_n$
(taking that $x^i$ are in the $n\times n$ 
irreducible representation)
\begin{equation}
 f=\sum_{l=0}^{n-1} \frac{1}{l!}\, f_{a_1\dots a_l}x^{a_1}\dots x^{a_l}
\end{equation}
tends in the limit $n\to \infty$  to the spherical harmonics 
expansion of the function $f$ on the sphere.

It seems apparent that this simple idea should be easy to 
implement to other maximally symmetric spaces defined as 
hyperspheres embedded in  higher-dimensional euclidean spaces,
by using suitable Lie groups and their Casimir relations.
It was applied in~\cite{Gazeau:2006hj,Jurman:2013ota} to obtain
2d and 4d fuzzy hyperboloids:
 in four dimensions however it is more difficult to
find the appropriate  metric structure. We shall in this
section find,  within the algebra 
of de~Sitter group $SO(1,4)$, two differential structures 
which give it a metric of the four-dimensional de~Sitter space.

Let us introduce the notation. We have 
ten generators  of the $SO(1,4)$ group
$M_{\alpha\beta}$, $\alpha,\beta = 0,1,\dots 4$;
the signature is $\,\eta_{\alpha\beta} = {\rm diag}(-++++)$. The
 commutation relations are 
\begin{equation}
 [M_{\alpha\beta}, M_{\gamma\delta}] = i(\eta_{\alpha\gamma} M_{\beta\delta}
 -\eta_{\alpha\delta} M_{\beta\gamma}-\eta_{\beta\gamma} M_{\alpha\delta}+
 \eta_{\beta\delta} M_{\alpha\gamma}).                     
\end{equation}
Using  $M_{\alpha\beta}\,$ one can define a vector $W_\alpha$ 
which is quadratic,
\begin{eqnarray}
& W_\alpha =\dfrac 18 \,\epsilon_{\alpha\beta\gamma\delta\eta}
 M^{\alpha\beta} M^{\delta\eta},                        \label{W}
\\[8pt]
& \ \ [M_{\alpha\beta}, W_\gamma] 
=i(\eta_{\alpha\gamma}W_\beta -\eta_{\beta\gamma}W_\alpha).                     \label{**}
\end{eqnarray}
The $SO(1,4)$ has  two Casimir operators:
\begin{equation}
 {\cal Q} = -\frac 12 \, M_{\alpha\beta} M^{\alpha\beta} , 
\qquad  \
{\cal W} =- W_\alpha W^\alpha  .
\end{equation}
In order to understand properties of the algebra with respect to
rotations in more detail, we adapt the notation to the $SO(3)$ 
subgroup  and denote the 3-vector indices by $i,j=1,2,3$.
We rename the generators,
\begin{equation}
  L_i =\frac 12 \,\epsilon_{ijk} M_{jk},\qquad  P_i = M_{i4}, 
\qquad
Q_i = M_{0i}, \qquad  R = M_{04}.
\end{equation}
The commutation relations of the $SO(1,4)$ are then
\begin{equation}
 \begin{array}{lll}
 ×[L_i,L_j]  =i\epsilon_{ijk}L_k,\qquad
&[L_i,P_j] =i\epsilon_{ijk}P_k,\qquad
&  [L_i,Q_j] =i\epsilon_{ijk}Q_k,
\\[6pt]
[P_i,L_j] = i\epsilon_{ijk}P_k,
&[P_i,P_j] = i\epsilon_{ijk}L_k,
&[P_i,Q_j] =i\delta_{ij} R,
\\[6pt] 
[Q_i,L_j] = i\epsilon_{ijk}Q_k,
&[Q_i,P_j] = - i\delta_{ij}R,
&[Q_i,Q_j] = - i\epsilon_{ijk}L_k,
\\[6pt]
[R, L_j ]=0=iQ_j, 
&[R, P_j] =iQ_j, &[R, Q_j] =iP_j .
\end{array}
\end{equation}

The algebra can be contracted in various ways. Rescaling 
$P_i \to P_i/\sqrt \Lambda$, $R\to  R/\sqrt \Lambda\,$,  
for $\Lambda \to 0$ we obtain the In\" on\" u-Wigner contraction 
to the Poincar\' e algebra, that is the flat limit of the de~Sitter
algebra: $R$ and $P_i$ become the generators of 
4-translations while $L_i$ and $Q_i$ generate 
the 3-rotations and boosts. In this limit
 $\,{\cal Q}$  and  $\,{\cal W}$ become the Casimir operators
of the Poincar\' e algebra: the square of  mass  and the
square of  Pauli-Lubanski vector.  Contraction $P_i \to \mu P_i$,
$Q_i \to \mu Q_i$, $R\to \mu^2 R\,$ for $\mu \to \infty$ gives
 phase space in three dimensions with rotations: $R$ becomes
a central element.
 
In  new notation components $W_\alpha$ are given by
\begin{align}
W_0 &= L_iP_i = P_i L_i, \qquad \  W_4 =-L_iQ_i =-Q_i L_i,
\\[4pt]
W_i &=RL_i +\epsilon_{ijk}Q_jP_k =RL_i -\epsilon_{ijk}P_jQ_k . \label{sm}
\end{align}
Commutation relations (\ref{**}) can be rewritten as
\begin{equation}
 \begin{array}{lll}
 ×[L_i,W_0] =0 ,   & [L_i,W_4] =0 ,        & [L_i,W_j] =i\epsilon_{ijk} W_k, 
\\[6pt]
[P_i,W_0] =0,  & [P_i,W_4] =-iW_i, \qquad&  [P_i,W_j] =i\delta_{ij}  W_4 ,
\\[6pt]
 [Q_i,W_0] =-iW_i,\qquad &  [Q_i,W_4] =0 ,  & [Q_i,W_j] =-i\delta_{ij} W_0 ,
\\[6pt]
 [R,W_0] =-iW_4,   \qquad&  [R,W_4] =-iW_0,    
& [R, W_j] =0,
\end{array}             \label{2} 
\end{equation}
and for the Casimir operators we obtain
\begin{eqnarray}
& - {\cal Q}
= -R^2 - Q_iQ_i + P_iP_i +L_i L_i ,            \label{quadr} \\[6pt]
&
-{\cal W} =  -(W_0)^2 +W_iW_i+(W_4)^2 = 
-(W_0)^2 -[W_0,Q_i]^2 -[W_0,R]^2 .               \label{quart}
\end{eqnarray}

Unitary irreducible representations of $SO(1,4)$ are 
in the notation of \cite{U,I,R} divided in four classes.
In the Class I, quadratic Casimir operator
 $\,{\cal Q}>0$ has continuous range
of values and the quartic Casimir operator is zero, $\,{\cal W}=0$.
 In the  Class II representations, $\,{\cal W}=0$ while $\,{\cal Q} $
is discrete, $\,{\cal Q} = -(n-1)(n+2) $,  $n=1,2,\dots$. 
The  Class III representations have continuous
$\,{\cal Q}$ and continuous $\,{\cal W}$, 
$\, {\cal W}= s(s+1){\cal Q} +(s-1) s(s+1) (s+2)$,
$s=1/2,1,3/2,\dots$ . Finally, in the Class IV representations both 
Casimir operators $\,{\cal Q}$ and $\,{\cal W}$ are discrete. Note 
that relation  (\ref{quart}) implies that if
$\, W_0=0$, then the quartic Casimir vanishes,  $\, {\cal W}=0$.

Let us discuss possible assignations of coordinates and momenta.
Casimir relation (\ref{quart})  directs us to take $W^\alpha$ 
as coordinates of a five-dimensional embedding space, and we shall 
adopt this identification.  The $SO(1,4)$ generators 
$M_{\alpha\beta}$ are as defined dimensionless, so we introduce
\begin{equation}
 x^\alpha = \ell \, W^\alpha ,
\end{equation}
where  constant $\ell$, of dimension of length, will be fixed later.
By definition noncommutative de~Sitter space ${\cal A}$ is the algebra 
generated by $x^\alpha$ in one of the unitary irreducible 
representations.  The quartic Casimir gives the value 
of the  cosmological constant,
\begin{align}
 \eta_{\alpha\beta}\,x^\alpha x^\beta =
- \ell^2{\cal W}   = \frac {3}{ \Lambda} \, .
\end{align} 
Coordinates $x^\alpha$  are quadratic in the group generators
 and therefore  they  do not close under commutation:
in terms of  decomposition~(\ref{sm}) we have
\begin{alignat}{2}
[x_0,x_i] &= i\epsilon_{ijk}\ell\, x_j P_k +i \ell\, L_ix_4,\qquad\ \
&[x_4,x_i] &= i\epsilon_{ijk} \ell\, Q_j x_k + i\ell\, L_i x_0 ,        \label{WWW}
\\[6pt]
[x_0,x_4] &= i \ell\, L_i x_i, 
&[x_i,x_j] &= i\epsilon_{ijk} \ell\, (Rx_k -P_k x_0  - iQ_k x_4). \label{WWW1}
\end{alignat}
These relations  are to some extent unsatisfactory because 
 their right-hand-sides  contain, besides coordinates,
the group generators: if only  $x^\alpha$  appeared, the 
interpretation would be much easier.  However this obstruction
 is not principal:  relations  (\ref{WWW}-\ref{WWW1})  
show only that calculations, if nontrivial, will not be simplified 
easily and will possibly depend on the representation.

Momenta have to fulfil stricter requirement,
they must close into an algebra which is at most quadratic.
In addition in order to interpret the frame  as
gravitational field, 
\begin{equation}
 g^{\alpha\beta}(x) = e^\alpha_A \, e^\beta_B\, \eta^{AB}  ,   \label{metric}
\end{equation}
we have to require that the frame elements depend on coordinates only, 
\begin{equation}
 [p_A, x^\alpha] =e^\alpha_A(x).                    \label{frameelements}
\end{equation}
It is thus clear that  $W^\alpha$  cannot be the momenta
 as on the fuzzy sphere. A natural choice would be to select 
momenta among the group generators. 
If we wish  to preserve the full de~Sitter
symmetry we shall choose as momenta  all $M_{\alpha\beta}$,
\begin{equation}
i p_A = \sqrt{ \Lambda}\, M_{\alpha\beta} ,              \label{pA}
\end{equation}
where the index $A$, $ A=1,\dots 10 $, denotes antisymmetric 
pairs $[\alpha\beta]$.  To get dimensions right we introduced 
in the last formula $\sqrt{ \Lambda}\,$; we 
systematically use convention $\hbar =1$.
The square root of the cosmological constant in (\ref{pA}) is
in fact  implied  by the frame formalism as, when momenta form  
a Lie group,
\begin{equation}
 [p_A, p_B] = C^D{}_{AB}\, p_D,
\end{equation}
 the curvature scalar is quadratic in the structure constants,
\begin{equation}
 R = \frac 14\, C^{ABD} C_{DAB}.                        \label{curvature}
\end{equation}
To be completely accurate we should have in fact put
\begin{equation}
 i p_A = \sqrt{ \zeta \Lambda}\, M_{\alpha\beta} ,     \label{zeta}
\end{equation}
 then the normalization of the scalar curvature,
$\, R=4\Lambda\,$, would have given $\,\zeta = 1/3$; we will
 however, as simpler, keep (\ref{pA}).
The constant $\ell$ we fix as
\begin{equation}
 \ell = \kbar \sqrt{ \Lambda},\qquad \ {\cal W}
=-\frac{3}{\kbar^2 \Lambda^2}  \, ,                          \label{CC}
\end{equation}
where $\kbar$ is the scale of noncommutativity of dimension 
length squared.  In consequence $\kbar$ enters only position  commutators,
\begin{alignat}{2}
[x_0,x_i] &= \kbar \,( \epsilon_{ijk}\, x_j p_{k+3} +p_ix_4) ,\qquad\
&[x_4,x_i] &=\kbar \, ( \epsilon_{ijk}  \, p_{j+6} \, x_k +  p_i x_0 ) ,     
\\[6pt]
[x_0,x_4] &= \kbar \, p_i x_i, 
&[x_i,x_j] &= \kbar\, \epsilon_{ijk}  \, (p_{10}x_k -p_{k+3} x_0  - p_{k+6} x_4).
\end{alignat}
Therefore the commutative limit is defined as $\kbar\to 0$.

The given choice of differential
structure might seem at the first sight unusual: we have 
spacetime  of $\, 5-1=4$ dimensions, with the tangent space of 
10 dimensions.  As we explained earlier, this comes formally 
with  noncommutativity of coordinates.
To understand the meaning of the introduced differential  $d$
we proceed to  the metric and the laplacian. 
Denoting
\begin{align}
ip_i =  \sqrt{ \Lambda}\, L_i,\quad ip_{i+3} = \sqrt{ \Lambda} \, 
P_i,\quad ip_{i+6} =  \sqrt{ \Lambda} \, Q_i,\quad ip_{10} =  \sqrt{\Lambda} \, R,
\quad i=1,2,3 ,                           \label{ABC}
\end{align}
and introducing a locally flat metric  $g^{AB}=\eta^{AB}$
 with signature $\, (++++++----)$ we obtain  the following expressions
for the  frame components,  $ \, e^\alpha_A =[p_A, x^\alpha]$
\begin{alignat}{5}
e^0_j &=0,
&e^0_{j+3} &=0
&e^0_{j+6} &= \sqrt{ \Lambda}\, x^j
&e^0_{10} &= \sqrt{ \Lambda}\, x^4 
\\[4pt]   
e^i_j &=-\epsilon^i{}_{jk}   \sqrt{\Lambda} \,x^k
\qquad &e^i_{j+3} &=\delta^i_j  \sqrt{  \Lambda} \, x^4
\qquad &e^i_{j+6} &= \delta^i_j   \sqrt{  \Lambda}\, x^0
\qquad &e^i_{10} &=0
\\[4pt]
e^4_j &=0,
&e^4_{j+3} &= -  \sqrt{ \Lambda}\, x^j
&e^4_{j+6} &= 0
&e^4_{10} &= \sqrt{ \Lambda}\,  x^0    .
\end{alignat}
From these expressions we can find
 differentials $\, dx^\alpha = e^\alpha_A \, \theta^A \, $
\begin{align}
dx^0 &=  \sqrt{  \Lambda}\, x^i \theta^{i+6} + \sqrt{ \Lambda}\, x^4 \theta^{10}, \\[4pt]
dx^i &= -\epsilon^{i}{}_{jk}  \sqrt{ \Lambda} \,x^k\theta^j + \sqrt{\Lambda} \, x^4 \theta^{j+3} +  \sqrt{ \Lambda}\, x^0 \theta^{j+6} ,
\\[4pt]
dx^4 & = - \sqrt{ \Lambda}\, x^i\theta^{i+3} + \sqrt{\Lambda}\, x^0\theta^{10} .
\end{align}
The spacetime components of the metric,
$\,  g^{\alpha\beta} = e^\alpha_A e^\beta_B \eta^{AB}, \  \alpha = 0,1,2,3,4 $,
are
\begin{equation}
 g^{\alpha\beta} =\Lambda \, \begin{pmatrix}
   -(x^i)^2-(x^4)^2 \  &  -x^i x^0 \ & -x^4x^0 \\
 -x^0 x^j  &\big(-(x^0)^2+(x^i)^2  +(x^4)^2\big)\, \delta^{ij}\, -x^j x^i
 & -  x^4 x^j\\ 
-  x^0 x^4  & - x^i x^4  & - (x^0)^2+(x^i)^2
 \end{pmatrix}  ;
\nonumber
\end{equation}
 this can be simplified to
\begin{equation}
 g^{\alpha\beta} =  3
\eta^{\alpha\beta} - \Lambda x^\beta x^\alpha .
\end{equation}
In the commutative limit $\,g^{\alpha\beta}$ is 
singular and reduces to the projector on the 4-dimensional 
de~Sitter space defined by (\ref{quart}); it projects
out the radius vector $x_\alpha$. When we  calculate the 
curvature which corresponds to the given frame, as
shown in the Appendix, we  obtain $\,  R=12\Lambda $;
 more generally, using rescaling (\ref{zeta}) we have
\begin{equation}
 R=12 \, \zeta \Lambda.                     \label{Rscalar1}
\end{equation}

Quadratic Casimir  operator $\, {\cal Q}$ is usually related 
to mass \cite{Garidi:2003ys,Gazeau:2006gq}. For a
scalar field $\Phi(x)$ the laplacian and the 
Klein-Gordon equation are written as
\begin{equation}
\Delta \Phi = - \eta^{AB} e_A e_B\Phi
=- \eta^{AB}[p_A,[p_B,\Phi]]=m^2\Phi ,          \label{lapl1}
\end{equation}
and clearly  mass corresponds to the value(s) of $\,{\cal Q}$ 
in the adjoint representation. In the contraction limit to 
the Poincar\' e group, generators $L_i$ and $Q_i$ become negligible 
so the mass (\ref{quadr}) reduces to
\begin{equation}
 -{\cal Q}\vert_{\Lambda\to 0}=-R^2 +(P_i)^2.
\end{equation}

It  possible to reduce the number of momenta and 
the dimensionality of the tangent space while keeping the metric
of $\, {\cal A}$ de~Sitter.  We introduce another set of momenta
\begin{equation}
 i\tilde p_0 =\sqrt\Lambda \, R,\qquad
 i\tilde p_i =\sqrt\Lambda \,({P_i+Q_i}), \quad i=1,2,3            \label{444}
\end{equation}
and denote the correspondingly defined differential by $\tilde d$;
 coordinates are  the same. The momentum commutators   are
\begin{equation}
  [\tilde p_0, \tilde p_i] = -\sqrt\Lambda \, \tilde p_i, \qquad\
[\tilde p_i, \tilde p_j] =0.
\end{equation}
Applying the algebra relations we get now for the frame elements
\begin{equation}
 \begin{array}{lll}
  \tilde e^0_0 =[\tilde p_0,x^0] = -x^4 ,  \quad &
\tilde e^i_0 =[\tilde p_0 , x^i] =0,\ \  &
\tilde e^4_0 =[\tilde p_0,x^4]=x^0,  \\[6pt]
\tilde e^0_j =[\tilde p_j,x^0] = x^j ,\ \  &
\tilde e^i_j =[\tilde p_j , x^i] =\delta_{ij} (x^0+x^4),\quad &
\tilde e^4_j =[\tilde p_j,x^4]= -x^j.
 \end{array}                                   \label{333}
\end{equation}
Therefore the coordinate differentials are given by
\begin{align}
\tilde dx^0 & = x^4 \tilde\theta^0 + x^j\tilde  \theta^j,
\\[4pt]
 \tilde d x^i & = (x^0+x^4) \, \tilde\theta^i  ,       \label{504}
\\[4pt]
\tilde dx^4 & =  x^0 \tilde\theta^0 - x^j\tilde  \theta^j,
\end{align}
and we easily recognize that  variable $\, x^0 +x^4\,$ should 
be introduced as a new coordinate,  time. Denoting
\begin{equation}
 \tau = -\log (x^0 +x^4)
\end{equation}
we find 
\begin{equation}
 d(x^0+x^4) = (x^0+x^4)\,\tilde\theta^0, \qquad 
\tilde\theta^0 = -d\tau.                                \label{505}
\end{equation}
The line element becomes in the classical limit that 
of  de~Sitter space,
\begin{equation}
 \tilde ds^2 = -(\tilde\theta^0)^2+ (\tilde\theta^i)^2 =
-\tilde d\tau^2 + e^{2\tau} \,\tilde dx^i \tilde dx^i.
\end{equation}
The remaining differential $\, \tilde dx^4\, $ is not independent:
its value follows from  the Casimir constraint
$\,  \tilde d\,  (x^\alpha x_\alpha) =0 $.
Assuming that the metric in frame components has the signature 
$(-+++)\, $ for the coordinate components  we obtain
\begin{equation}
\tilde g^{\alpha\beta} =\begin{pmatrix}
  -(x^4)^2 +(x^i)^2 \  & x^i (x^4+x^0) \ & -x^4x^0 - (x^i)^2\\[2pt]
x^i (x^4+x^0) & \delta^{ij} \, (x^4+x^0)^2 & - x^i (x^4+x^0) \\[2pt] 
-x^4 x^0 - (x^i)^2 & - x^i (x^4+x^0)  & - (x^0)^2+(x^i)^2
 \end{pmatrix}  .
\end{equation}
Again this is a projector to the hypersphere which projects 
out $x_\alpha$.

We achieved to obtain four-dimensional tangent space.
 The Laplace operator induced by $\tilde d$ is given by
\begin{equation}
\tilde  \Delta \Phi =\Lambda\, [R,[R,\Phi]] -\Lambda\, [P_i+Q_i,[P_i+Q_i,\Phi]]
=\tilde m^2 \Phi,                                 \label{lapl2}
\end{equation}
and although it is not invariant under the full de~Sitter group,
it has the correct Poincar\' e limit under the contraction,
\begin{equation}
 -{\tilde m^2}\vert_{\Lambda\to 0}=-R^2 +(P_i)^2 .
\end{equation}
The laplacian $\tilde \Delta$ invariant under the 3-rotations and boosts,
\begin{equation}
 [L_j, -R^2 +(P_i)^2]=0,\qquad    [Q_j, -R^2 +(P_i)^2]=0.
\end{equation}
As shown in the Appendix, the scalar curvature related 
to $\tilde d$ is constant,
\begin{equation}
 \tilde R = \frac 34\, \Lambda,                \label{Rscalar2}
\end{equation}
and if we wish to have the usual value $4\Lambda\,$ 
for the scalar curvature we need to
redefine the momenta, $\, \tilde p_\alpha\to \tilde \zeta\, 
\tilde p_\alpha\,$ with  $\, \tilde \zeta =16/3$. The fact 
that $\tilde \Delta$ is not invariant under the full de~Sitter group 
indicates that the metric is not invariant too: indeed 
 this is the case and it can be confirmed by an
analysis of the Killing equations which is done in the Appendix.

\section{Noncommutative de~Sitter space, II}

To construct a four-dimensional de~Sitter space we 
started  in the previous section
from the algebra of de~Sitter group $SO(1,4)$. Within this
algebra we  identified coordinates,
momenta and the moving frame consistently with 
rules imposed by the frame formalism: as result we found
two natural realizations of fuzzy de~Sitter space. 
A drawback of this procedure is the existence of a number of operators
which have no direct physical interpretation, while they cannot be 
avoided in calculations: one would prefer to have an algebra which 
is as small as possible, minimal.

 One method to search for such an algebra is
to try to construct it: first, to assume that it exists 
by making an Ansatz for the commutation relations; one needs a
further Ansatz for the moving frame, the one that gives the required
form of the metric. Using these two premises one 
 solves the constraints: the Jacobi, the frame and the
 compatibility equations. Such programme can be systematically 
done only approximately, in linear order in noncommutativity 
(otherwise, in the course of calculations one would have to use the 
multiplication rules  which one is looking for) and  
as  result one  obtains a
noncommutative spacetime, that is an algebra and a calculus.
Following this approach we have made in our previous papers  a 
survey of noncommutative algebras generated by four, five and six
elements, \cite{Buric:2014ika,Buric:2009zz,Buric:2008th}. 
The original motivation was in part to use interior 
derivatives, thereby  decreasing the dimension of phase space. 
The summary of the results dimensionwise is  as follows.
The lowest-dimensional spacetime which can be constructed has 
four generators (the corresponding phase space has five), 
but the metric is necessarily nonstatic: it is in fact 
unique. When we add one more generator, that is, within 
the set of five-dimensional position  algebras, 
constraints get considerably relaxed 
and we obtain a large family of static spherically symmetric 
noncommutative geometries.

We wish here to generalize the frame Ansatz and the constraints 
of \cite{Buric:2014ika} in order to obtain noncommutative cosmological  
spacetimes of  the Friedmann-Robertson-Walker type. Let us 
briefly introduce the notation and write down the equations;
for all technical details  we refer to \cite{Buric:2014ika} as
calculations are to some extent analogous. Coordinates  are denoted by 
\begin{equation}
 x^0=t,\qquad  x^a =\rho \xi^a, \qquad x^4 =r, \ \qquad a=1,2,3.
\end{equation}
Normalized vector $\xi^a$ describes two angular variables, 
polar and azimuthal angle;  radial coordinate 
(measuring either the area of the sphere or the radial distance)
 is  $r$, time is $t$. The $x^a$ satisfy
\begin{equation}
x_a x^a =\rho^2,                               \label{rho2}
\end{equation}
therefore the additional fifth coordinate $\rho $ is 
analogous to a radius but of an additional, extra dimension.
The $\xi^a$ are the generators of the $SO(3)$ algebra in the
irreducible $n\times n$ representation: momenta $p_a$ are taken to be
 proportional to $\xi^a$, as on the fuzzy sphere.

We assume that the form of position algebra ${\cal A}$ is 
\begin{eqnarray}
&\ [\xi^a,\xi^b] = \dfrac{2i}{n}\, \epsilon^{abc}\, \xi_c, \qquad\quad
[\xi^a,\rho]= [\xi^a,r]=[\xi^a,t]=0     ,        \label{1aa}
\\[8pt]
&[\rho,t] = \kb J^0 \rho,   \qquad\ \
[r,t] = \kb J ,                \qquad   \ \
[\rho,r] = \kb J^4 \rho    ,
\end{eqnarray}
that is,  a tensor product $so(3)\otimes {\cal A}^\prime\,$
where  ${\cal A}^\prime\, $ is the algebra generated by $r$, $t$ 
and $\rho$. Differential calculus
mixes the two subalgebras: the Ansatz for the frame  is given by
\begin{equation}
\begin{array}{ll}
 \theta^a = -{h}{\rho}^{-1}\, \epsilon^{a}{}_{bc}\, x^b dx^c 
+\rho^{-2} \,x_b \theta^b x^a ,
       \qquad \  & 
dx^a =({h\rho})^{-1}\, \epsilon^{a}{}_{bc}\, x^b \theta^c  , \\[6pt]
\theta^4 = g\,dr    , &    dr = g^{-1} \theta^4      ,                   \\[6pt]
 \theta^0 = f dt + k x^a\theta^a ,  &          dt = f^{-1} \theta^0 -kf^{-1}  x^a\theta^a    .
\end{array}      \label{d1a} 
\end{equation}
This Ansatz implies, as on the fuzzy sphere, 
\begin{equation}
 \rho d\rho + d\rho\, \rho =0 ,
\end{equation}
that is,  $\rho $ is in the commutative limit a constant.
To impose spherical symmetry  we assume that
$f$, $g$, $h$ and $k$, as well as $J$, $J_0$ and $J_4$
 are functions of $\rho$, $r$ and $t$ only. 
We thus have seven unknown functions 
which are to be determined from the constraint equations.
We will consider here only the case  $\, k=0$ which gives  the 
usual diagonal form of the metric; $k\neq 0$  allows to 
extend the construction to the generalized Taub-NUT spaces.  

Consistency of the frame with the algebra, the Jacobi and 
the frame constraints give the following set of equations:
\begin{equation}  
 \begin{array}{ll}
\dot J^4=0  ,\qquad  \qquad J^{0 \prime} =0,       &
 h^\prime J^4 + \dot h J^0 =0   ,
 \\[6pt]
J^{4\prime} + {g}^{-1}  (g^\prime J^4 +\dot g J^0) =0,
 &  (h +\rho \p_\rho h)  J^4
- \dot h J =0  , \\[6pt]
J^{0\prime} +{f}^{-1} (f^\prime J^4 
+\dot  f J^0)=0,
&  {( h+\rho \p_\rho h)  J^0
+h ^\prime J = f^{-1}\, k \, h^2\rho^2 ,} 
\\[6pt]
 \dot J + f^{-1} ( \dot f J 
-\rho \p_\rho f J^4)=0 ,
\quad &   J^\prime + g^{-1} ( g^\prime J 
+ \rho \p_\rho g J^0 ) =0 ,
\\ [6pt]
 k^\prime J^4 + \dot k J^0 =0  ,
& (k +\rho \p_\rho k)  J^4
- \dot  k J =0  ,                               \label{E3}
 \end{array}
\end{equation}
where $\dot f =\p_t f$ and $f^\prime =\p_r f$. Equations are 
coupled and nonlinear, and thus relatively complicated;
but we do not need to solve them in full generality because of
the diffeomorphism invariance of the formalism
which can be used to fix some of the variables.
 We choose  the radial coordinate such 
that the area of sphere be $4\pi r^2$, that is we put $\, h=r$.
We assume further that the metric has
 the Schwarzschild form, $\ fg=1$. In the static
 case, when no function depends on time, we find a solution
\begin{eqnarray}
&& J^4=0          , \qquad J^0=J^0(\rho),   \qquad     J=- r J_0(\rho) ,                                                                       \label{j4a}
\\[6pt] 
&& h=r,  \qquad   \quad  f=\dfrac 1g =\gamma r \, F(\rho r) ,    \label{form}
\end{eqnarray}
function $F$ can be arbitrary function of its argument. The 
corresponding metric in the commutative limit has the line element
\begin{equation}
 ds^2 =-f^2 dt^2 +\frac{1}{f^2} \, dr^2 + r^2 d\Omega.     \label{48}
\end{equation}
Though (\ref{48}) is restricted in its form, 
it gives a large family of solutions. We have seen that in the 
 classical limit   $\rho$ becomes a  constant: therefore  taking 
for example 
\begin{equation}
 F(\rho r) = \sqrt{ \frac {1-\rho^2 r^2}{\rho^2 r^2} }  \,
\end{equation}
and identifying $\rho$ with the cosmological constant,
\begin{equation}
 \rho^2=\gamma^2 =\frac {\Lambda}{3} \, ,           \label{rhoL}
\end{equation}
  we obtain the de~Sitter metric in static coordinates. 
As a consequence, we find that  the cosmological constant is quantized.

Noncommutative de~Sitter space ${\cal A}\, $ is in this version 
 a relatively small algebra, generated by five
elements; it is in contrast to the spacetime found in previous 
section closed. But to obtain  phase space we have to extend it by one
generator, $p_4$. Namely from (\ref{d1a}) and our solution
we see that momentum $p_4$ has to satisfy
\begin{equation}
 [p_4, t]=0,\qquad [p_4, x^a] =0,\qquad [p_4,\rho r]=0,   \label{pP}
\end{equation}
and therefore it does not belong to $\,{\cal A}$. These relations 
can be solved, not uniquely,  within a larger algebra.

In order to find within this same framework the
Friedmann-Roberstson-Walker type 
 solutions,  we should include the dependence on time.
It is necessary also to take $ \,  J^4\neq 0$. 
Again, instead of attempting to find the most general solution
 we seek for a particular one, of the form
\begin{equation}
 f=1,   \qquad g=a(t) \, G(r,\rho), \qquad h=a(t)\,  r.
\end{equation}
We obtain the following set of equations:
\begin{eqnarray}
&& \p_t J^0 =0, \qquad \qquad \qquad\  \quad\ \p_t J^4 =0, \\[6pt]
&& \p_t J=0 ,  \qquad \qquad  \qquad \qquad
 \p_r J^0 =0,
\\[6pt] 
&& aJ^4 +\dot a J^0 =0,\qquad   \qquad\quad aJ^4 -\dot a J =0,
 \\[6pt] 
&&
\p_r J J^4 -\p_r J_4 J -\rho \p_\rho J^4 J^0 +\rho \p_\rho J^0 J^4 =0.
\end{eqnarray}
These equations have a solution similar to (\ref{j4a})
\begin{equation}
 J^0 = J^0(\rho),\qquad J=-rJ^0, \qquad J^4 = -\frac{\dot a}{a}\, r J^0.
\end{equation}
The remaining constraint gives 
 $\, G(r,\rho) = \rho $, so we find for the frame
\begin{equation}
 f =1,\qquad g = a(t) \, \rho, \qquad h= a(t)\, r.
\end{equation}
The scale factor $a(t)$ can be arbitrary;
the limiting classical metric is given by
\begin{equation}
ds^2 = -dt^2 +a^2(t) \big( \rho^2 dr^2 + r^2 d\Omega \big),
\end{equation}
where $\rho$ is as before constant so we can
 choose $\rho =1$. Taking for example the exponential 
function,
\begin{equation}
a(t) = \exp {\sqrt{\frac \Lambda 3}  \, t}  \, ,
\end{equation}
we obtain the de~Sitter space in the FRW form. In general, 
we find a family of fuzzy Friedmann-Robertson-Walker geometries.

\section{Conclusions}

We constructed in this paper, using the noncommutative
 frame formalism, essentially two different versions of 
noncommutative four-dimensional de~Sitter space. Let us review 
and compare them.

De~Sitter spaces constructed in Section 2 are based 
on the algebra of  de~Sitter group $SO(1,4)$. 
Coordinates which generate spacetime $\,{\cal A}$ are 
proportional to  5-vectors $W^\alpha$ (\ref{W}),
so in the unitary irreducible representations of the group,
 Casimir relation (\ref{quart})  defines  four-dimensional
de~Sitter space as an embedding in  five dimensions.
The value of the quartic  Casimir operator is 
related to the cosmological constant $\Lambda$ and  
noncommutativity scale $\kbar$, (\ref{CC}); whether $\Lambda$ has 
discrete or continuous spectrum depends on the representation.
Spacetime $\,{\cal A}$ has somewhat unpleasant property
that coordinates do not close under commutation. But as it is
explained in the text, this aspect is not essential: what is really
necessary (and sufficient) for physical interpretation is that
commutators with momenta (\ref{frameelements}), which
give the frame and the metric, can be written  
in terms of coordinates solely. Within $\,{\cal A}$ 
we formulate two differental calculi both based on 
the generators of the de~Sitter group, $M_{\alpha\beta}$:
 of course, other differential structures can be constructed
 but they give other geometries.
The first calculus has as  momenta all group generators and
 therefore the corresponding tangent space is 10-dimensional.
The second calculus has four momenta and the  tangent space which
is 4-dimensional. In both cases metric in coordinate components
(\ref{metric}) is a 5$\times$5 matrix which  projects to 
4-dimensional de~Sitter space $\,{\cal A}$; also,  the scalar 
curvature is in both cases constant. 
The two calculi have different laplacians: laplacian $\Delta$
is related to the quadratic Casimir of the $SO(1,4)$ which is
in field theories on the commutative de~Sitter space usually 
related to mass. Second  laplacian $\tilde\Delta$  is not invariant 
under the full de~Sitter symmetry, but it is invariant under 
the Lorentz subgroup. We therefore obtain two different Klein-Gordon 
equations: in both cases the In\" on\" u-Wigner contraction gives 
the usual mass.

Section 3 contains two more versions of fuzzy de~Sitter space,
 one in the static and  another in the FRW coordinates: we can
obtain in fact 
 noncommutative versions of all Friedmann-Robertson-Walker 
spacetimes. The strategy of the calculation is in 
Section 3 inverted: instead of fixing the 
algebra in advance we search for it, and therefore position space 
$\, {\cal A}$ which we obtain is always, by definition, closed. 
However it is not minimal: it  has $6-1=5$
generators, (\ref{rho2}). Although the
additional variable $\rho $ is not a Casimir operator,
 it is constant in the sense that its differential 
is zero,  $d(\rho^2) =0$. We interpret $\rho$ as a size
of the additional dimension: classically
its value is related, in the cosmological case,
  to $\Lambda$ and $\kbar$, (\ref{rhoL}).
We have not solved, within this approach, for the full phase 
space algebra and therefore the laplacians are missing:
equations for $p_\alpha$ are  however given, (\ref{pP}).
One of the problems with searching for algebras which
satisfy equations like (\ref{pP}) is that there are `too
many' zeroes on the right-hand-sides, and consequently
too many extensions are possible: it is much easier to 
solve the Jacobi identities when they  contain nonsingular
expressions. This presents also the main technical difference
 between the two approaches presented. On the one
hand, the {\it a priori} choice of the algebra of the $SO(1,4)$ 
group done in Section 2 might seem quite arbitrary: in principle,
 one would prefer to have a set of  equations and to find a unique 
solution to them. But on the other hand it is hard to imagine 
that, starting from  relations (\ref{444}), one would arrive at 
the de~Sitter algebra: there are many more `equally good' algebras
which contain these equations.

 In fact, the algebra of the $SO(1,4)$ (or the $SO(2,3)$) group  
seems to be the ideal framework for noncommutative four-dimensional 
geometries. It is `big enough'  to be a phase space:
 the number of its generators is $10-2=8$.
 Also, the existence of three independent 3-vectors 
and one scalar gives enough room to construct rotationally 
invariant spaces: this is seemingly impossible 
within the algebra of $SO(1,3)$. 
Among various versions of fuzzy de~Sitter space  here analyzed, 
the one based on the de~Sitter algebra with four-dimensional 
tangent space is perhaps the most appealing, because of its 
dimension. It has an additional merit: it can be generalized 
to arbitrary even-dimensional space. Namely, 
 starting from the $SO(1,n)$ group and its generators
$M_{\alpha\beta}$, $\,\alpha,\beta = 0,1,\dots n$,  for even $n$
there is a vector  
\begin{equation}
W^\alpha =\epsilon^{\alpha\alpha_1 \dots\alpha_n}M_{\alpha_1\alpha_2} \dots
M_{\alpha_{n-1},\alpha_n} .
\end{equation}
The square of $\, W^\alpha$ is the Casimir operator of the highest 
rank in  $SO(1,n)$. The corresponding Casimir relation
defines an embedding of the $n$-dimensional hypersphere
in a ($n$+1)-dimensional  space. By analogy with the 
four-dimensional case, the choice of
\begin{equation}
i\tilde p_0 = M_{0n}, \qquad \ i \tilde p_i = M_{0i}+M_{in}, 
\quad i=1,2,\dots n-1  ,
\end{equation}
as momenta gives the $n$-dimensional tangent space. 
We find therefore  $n$-dimensional fuzzy de~Sitter spacetime
 in coordinates which give the flat sections, together with 
the corresponding time-dependent metric.

The other fuzzy de~Sitter space based on the $SO(1,4)$
group, with ten-dimensional tangent space, is interesting
because it preserves the full de~Sitter symmetry. Somewhat 
counterintuitive is a large discrepancy of
dimensionalities of spacetime and  tangent
space,  4 and 10; though as we discussed earlier,
 equality of these two numbers 
is hardly to be expected.  There are many physical  
arguments which support the intuition that at small 
scales coordinates are  noncommuting operators. 
 One can develop similar intuition regarding the linear 
space of differentials: namely, if differentials are in
some way related  to the fluctuations of coordinates, 
there is no reason why  they cannot have `more dimensions' 
than coordinates themselves: afterall, fluctuations depend
on the states of the system too.  Similarly, one can
argue that a coordinate restricted to the `subspace
of a noncommutative manifold' can fluctuate, besides 
tangent, in orthogonal directions too. 
The difference in dimensions which we have here 
and in other cases \cite{fs} gives
completely  different counting of the degrees of freedom
of fields on a noncommutative space and it can potentially give
very interesting  models.

In any case,   both of the $SO(1,4)$-based models
of fuzzy de~Sitter space can be very useful.
Their  unitary irreducible representations are known, and that
gives many possibilities for further investigation: one can 
for example discuss spectra of coordinates, coherent states, 
field theories  etc. An equally important task would be
 to obtain perturbations as it was 
done in \cite{Buric:2006di} for perturbations of the flat space.
Finally, fuzzy de~Sitter space proves again
that the noncommutative frame formalism is
 a good and systematic method to describe noncommutative 
geometries, but also that we need more work to explore 
all its properties in order to use it properly.

\vskip0.5cm
\begin{large}
{\bf Acknowledgement}
\end{large}
This work was supported by the Serbian Ministry of Education, Science
and Technological Development Grant ON171031. The authors worked on
the subject while visiting  AEI in Berlin and they would like to
thank  Hermann~Nicolai for his hospitality.

%
%

\section{Appendix}

We give technical details related to the 
calculations and results given in Section 2; 
for a more  rigorous introduction we refer to \cite{book}.
Both differential calculi considered in 
Section 2 have  momenta which close into a Lie algebra.
In this case properties of the space of $p$-forms simplify:
one can show  for example that the frame forms anticommute, 
$\, \theta^A \theta^B + \theta^B \theta^A=0$, which is not
the case in general. Expression for the scalar
curvature is also very simple, (\ref{curvature}), so in
order to calculate it we only need to determine the structure 
constants.

If we take as momenta all de~Sitter group generators $p_A$, using 
enumeration (\ref{ABC}) we find that nonvanishing $C_{ABC}$ 
are
\begin{eqnarray}
  C_{ijk}=\epsilon_{ijk},\quad C_{i,j+3,k+3}=\epsilon_{ijk},
\quad C_{i,j+6,k+6}= - \epsilon_{ijk} ,\quad
 C_{i+3,j+6,10}= -\delta_{ij}.              \label{C10}
\end{eqnarray}
Nonzero are also all permutations of the structure 
constants with the given indices,  and  $C_{ABC}$ 
are completely antisymmetric. Using (\ref{C10})
one can easily obtain  (\ref{Rscalar1}).
For the second set of momenta $\tilde p_\alpha$
the nonvanishing structure constants are
\begin{equation}
 \tilde C_{i0j}=-  \tilde C_{ij0} =\delta_{ij}.
\end{equation}
These constants are neither cyclic nor fully antisymmetric;
they give however a constant curvature scalar,  (\ref{Rscalar2}).

Antisymmetry of the structure constants is important when we 
calculate the Lie derivative of the metric. Lie derivative, 
with its usual properties, can in general be defined on  
noncommutative spaces \cite{book}; we will use it here to check
whether derivations defined by the group generators are also 
the Killing vectors. The interior product of a vector field $X$ 
and a 1-form $\eta$ is given by
\begin{equation}
 i_X\eta = \eta(X).
\end{equation}
The action of Lie derivative ${\cal L}_X$ can be
extended, from the action on functions,
\begin{equation}
 {\cal L}_Xf = Xf ,
\end{equation}
 to the action on 1-forms and their products by linearity, Leibniz rule 
and the formula
\begin{equation}
  {\cal L}_X = i_X d + d\, i_X.                \label{Lied}
\end{equation}
In particular, using the fact that the frame components of the
metric are constants,
\begin{equation}
 g(\theta^A\otimes\theta^B) = \eta^{AB},
\end{equation}
we  find how $\,{\cal L}_X$ acts on the metric:
\begin{equation}
( {\cal L}_X g )(\theta^A\otimes\theta^B)=
-g({\cal L}_X \theta^A\otimes\theta^B ) -
g(\theta^A\otimes {\cal L}_X \theta^B).               \label{499}
\end{equation}
That is, in order to check whether vector
field $X$ is a Killing vector,
\begin{equation}
 {\cal L}_X g =0,
\end{equation}
  we need only to determine  $\,{\cal L}_X \theta^A $
and apply (\ref{499}). Using formula (\ref{Lied}) we can 
easily find the Lie derivatives corresponding to frame vectors $e_A$,
\begin{equation}
 \mathcal{ L}_{e_A}\,\theta^B = - C^B{}_{AD}\theta^D.        \label{500}
\end{equation}
We thus have, in the case of the first calculus,
\begin{equation}
( {\cal L}_{e_A} g )(\theta^B\otimes\theta^D)=
C^B{}_A{}^D+C^D{}_A{}^B =0,
\end{equation}
that is, all derivations $e_A$ are the Killing vectors and 
spacetime has the de~Sitter group as its symmetry.
For the second calculus this is not the case: 
we have for example
\begin{equation}
( {\cal L}_{\tilde e_0} g )(\theta^i\otimes\theta^j)=2 \delta^{ij}, 
\qquad 
( {\cal L}_{\tilde e_i} g )(\theta^0\otimes\theta^j)= \delta^{ij}, 
\end{equation}
so neither $\tilde e_0\,$ nor $\tilde e_i\,$ are Killing vectors.
 Rotations however are a symmetry of the metric,
\begin{equation}
  {\cal L}_{X_i} g =0  ,
\end{equation}
where $\, X_if =[L_i,f]$. This can be confirmed by 
extrapolating formula (\ref{500}) to the generators of rotations 
using (\ref{Lied}) and expansions (\ref{504}), (\ref{505}).

\end{document}